# Random Birth-and-Death Networks

Xiaojun Zhang[1*], Zheng He[2], Lez Rayman-Bacchus[3]

[1] *School of Mathematical Sciences,* [2]*School of Management and Economics,*

*University of Electronic Science and Technology of China, Chengdu, P. R. China, 611731*

[3] *University of Winchester, Winchester, SO22 5HT, UK*

**Abstract**

In this paper, a baseline model termed as random birth-and-death network model (RBDN) is considered, in which at each time step, a new node is added into the network with probability $p$ ($0< p <1$) connect it with $m$ old nodes uniformly, or an existing node is deleted from the network with probability $q=1-p$. This model allows for fluctuations in size, which may reach many different disciplines in physics, ecology and economics. The purpose of this study is to develop the RBDN model and explore its basic statistical properties. For different $p$, we first discuss the network size of RBDN. And then combining the stochastic process rules (SPR) based Markov chain method and the probability generating function method, we provide the exact solutions of the degree distributions. Finally, the characteristics of the tail of the degree distributions are explored after simulation verification. Our results show that the tail of the degree distribution for RBDN exhibits a Poisson tail in the case of $0<p\leq 1/2$ and an exponential tail as $p$ approaches to 1.

*Corresponding author: Xiaojun Zhang Email address: sczhxj@uestc.edu.cn

# Random Birth-and-Death Networks

## 1. Introduction

In the real world, most networks like the World Wide Web [1-3], friendship networks [4-7], communications networks [8-10], and food-web [11-13] are evolving with frequent node births (additions) and deaths (deletions), in which each agent is intelligent and has its own life cycle. Recently, these addition-deletion networks have caught much attention [14-24]. Various models have been developed to describe different evolving processes. Typically, Moore et al. [18] discussed addition-deletion networks in which at each unit of time one node is added, and with probability $p$ a randomly chosen node is deleted; Sarshar and Roychowdhury [16] considered an *ac hoc* network where new nodes joining the network make links preferentially and existing nodes are uniformly deleted at a constant rate; Slater et al. [17] considered a growing tree model allowing the possibility of death; Saldaña [20] discussed growing random networks with addition and deletion of nodes based on a differential mass balance equation; Ben-Naim and Krapivsky [21] discussed growing random networks in which at each time step, with rate $\alpha\,(\alpha \geq 1)$, a node is added to the network and with rate 1, a randomly selected node is deleted with its parent node inheriting the links of its immediate descendants. Although all these models deal with the deletion of nodes in the network, they at the same time add nodes at each time step to keep the network size growing or unchanged. In fact, the network sizes of many evolving networks like the Internet or social networks often fluctuate due to the appearance of new computers or persons respectively at one time or the disappearance of previously existing nodes at another time. This dynamic process may lead to the decreasing of the network size, reflecting that shrinking networks are also popular in reality.

In this paper, a more general random birth-and-death network (RBDN) is discussed, in which at each time step, a new node is added to the network with probability $p\,(0 < p < 1)$ or an existing node is deleted from the network with probability $q = 1 - p$. This RBDN model is simple enough to constitute a baseline model for many studies in dynamical networks, which reach many different disciplines in physics, ecology and economics. Meanwhile, as its main strength, it allows for fluctuations in size, granting it additional realism. Potentially, this model may allow scientists to decouple effects caused by dynamism itself from effects caused by the dynamic mechanism used.

Thus the aim of this study is to develop the RBDN model and explore its basic statistical properties.

This paper is organized as follows. Section 2 introduces the RBDN model. Section 3 discusses the network sizes of RBDN for different $p$ and Section 4 calculates the degree distributions of RBDN for different $p$. In Section 5, we first use computer simulations to verify our results in Section 4 and then explore the tail characteristics of the degree distributions for RBDN. Section 6 concludes the paper and proposes further direction.

## 2. RBDN Model

As noted above, previous models of addition-deletion networks keep the size of the network growing or unchanged, and all ignore that real-world networks may either expand or contract over time. Here we introduce a more general model termed Random Birth-and-Death Network (RBDN) model as follows.

(i) The initial network is an isolated node;

(ii) At each unit of time, add a new node to the network with probability $p$ $(0 < p < 1)$ and connect it with $m$ old nodes uniformly, or randomly delete a node from the network with probability $q = 1 - p$.

Note:

(a) During the evolving process, there exists a network size lower bound $n_0$, namely, if the number of nodes in the network is $n_0$ at time $t$, then at time $t+1$, we add a new node to the network with probability $p$ and randomly connect it to $m$ old nodes in the network, or keep it unchanged with probability $q$. To simplicity, here we define $n_0 = 1$. Indeed, in our study, different $n_0$ will not affect the distribution type and its properties.

(b) If at time $t$, a new node is added to the network and the network size is less than $m$, then the new node is connected to all old nodes.

(c) If at time $t$, a node is deleted, then all the edges incident to the removed node are also removed from the network, thus the degree of its neighbors decreases by one.

Since $p = 1$ corresponds to the pure addition network (which has been studied in [8]), and $p = 0$ corresponds to the pure shrinking network which is a single node, here we assume $0 < p < 1$. In the following sections, we will investigate some important properties of RBDN.

## 3. Network Size

Network size is fundamental to all networks since it may affect many others properties including the degree distribution, average path length and clustering coefficient [21]. Thus in this paper, we first discuss the network size of RBDN. For a growing network, it is obvious that the network size will be infinite at the limit $t \to \infty$. However, for RBDN, the network size is determined by different $p$.

Let $N(t)$ be the number of nodes in RBDN at time $t$ and $N(0)=1$. Let

$$\Pi_N(n) = \lim_{t \to +\infty} P\{N(t)=n\} \quad n \ge 1 \tag{1}$$

be the probability distribution of the network size of RBDN in the limit $t \to \infty$. Let $Q=(q_{i,j})$ be the one-step transform probability matrix of $\{N(t), t \ge 0\}$, where

$$q_{i,j} = P\{N(t+1)=j \mid N(t)=i\} \tag{2}$$

and $Q$ satisfies

$$q_{i,j} = \begin{cases} q & i,j=1 \\ q & j=i-1, i \ge 2 \\ p & j=i+1, i \ge 1 \end{cases} \tag{3}$$

Since the one-step transform probability matrix $Q$ is independent of time, $\{N(t), t \ge 0\}$ is a Markov chain with stationary transition probabilities. Also because $\{N(t), t \ge 0\}$ is a one-dimension random walk with a left bound 1 [25], we may draw the following conclusions easily.

(a) $0 < p < \frac{1}{2}$

$$\Pi_N(n) = \frac{q-p}{q}\left(\frac{p}{q}\right)^{n-1} \quad n \ge 1 \tag{4}$$

(b) $\frac{1}{2} < p < 1$

$$\Pi_N(n) = \begin{cases} 1 & n = +\infty \\ 0 & else \end{cases} \tag{5}$$

(c) $p = \frac{1}{2}$

$$\Pi_N(n) = 0 \quad n \ge 1 \tag{6}$$

As shown in Eqs. (4)-(6), the network size of RBDN follows geometric distributions in the case of $0 < p < \frac{1}{2}$. In the case of $\frac{1}{2} < p < 1$, the network size tends to be infinite as $t \to \infty$. For the special case $p = \frac{1}{2}$, the size of the network can uniformly be any positive integer.

## 4. Degree Distribution

Many methods have been used to calculate the degree distributions of evolving networks such as the mean-field method [26], the rate-equation method [27], the master-equation approach [28], the renormalization group [29], and the Markov chain method [24,30,31]. In this section, the stochastic process rules (SPR) based Markov chain method [24] is employed since it is effective for both node addition and deletion in the evolving network. In addition, we enhance the SPR method by using a probability generating function approach to solve the degree distribution equations for different $p$.

For the SPR method, there is an infinite number of isolated nodes at time $t = 0$. At each time step, the added nodes are provided by the other networks that have the same topologic structure, and the deleted nodes will construct new networks. Thus SPR method keeps the number of nodes unchanged at any time and maintains the topological structures and statistical characteristics [24]. Here for any node $v$, we use $(n,k)$ to describe the state of node $v$, where $n$ is the number of nodes in the network that contains $v$, and $k$ is the degree of node $v$. Let $NK(t)$ denote the state of node $v$ at time $t$ and $\tilde{P}(t)$ be the probability matrix of $NK(t)$, i.e.

$$\tilde{P}_{(n,k)}(t) = P\{NK(t) = (n,k)\} \tag{7}$$

Let $\boldsymbol{P}$ be the one-step transition probability matrix of $\{NK(t), t \geq 0\}$,

$$\boldsymbol{P} = \left(p_{(n_1,k_1),(n_2,k_2)}\right) \tag{8}$$

using SPR approach, $\boldsymbol{P}$ can be obtained（see the Appendix）, satisfying

$$\tilde{P}(t+1) = \tilde{P}(t) \cdot \boldsymbol{P} \tag{9}$$

Extending Eq.(9), we can deduce the state transformation equations of $NK(t)$ as follows:

$$
\begin{cases}
\tilde{P}_{(1,0)}(t+1) = q\tilde{P}_{(1,0)}(t) + q\tilde{P}_{(2,0)}(t) + q\tilde{P}_{(2,1)}(t) \\
2\tilde{P}_{(2,0)}(t+1) = 2q\tilde{P}_{(3,0)}(t) + q\tilde{P}_{(3,1)}(t) \\
\qquad\qquad \vdots \\
(m+1)\tilde{P}_{(m+1,0)}(t+1) = (m+1)q\tilde{P}_{(m+2,0)}(t) + q\tilde{P}_{(m+2,1)}(t) \\
(m+2)\tilde{P}_{(m+2,0)}(t+1) = (m+2)q\tilde{P}_{(m+3,0)}(t) + q\tilde{P}_{(m+3,1)}(t) + p\tilde{P}_{(m+1,0)}(t) \\
\qquad\qquad \vdots \\
n\tilde{P}_{(n,0)}(t+1) = nq\tilde{P}_{(n+1,0)}(t) + q\tilde{P}_{(n+1,1)}(t) + (n-m-1)p\tilde{P}_{(n-1,0)}(t) \\
\qquad\qquad \vdots
\end{cases} \tag{10}
$$

$$
\begin{cases}
2\tilde{P}_{(2,1)}(t+1) = q\tilde{P}_{(3,1)}(t) + 2q\tilde{P}_{(3,2)}(t) + p\tilde{P}_{(1,0)}(t) + p\tilde{P}_{(1,0)}(t) \\
3\tilde{P}_{(3,1)}(t+1) = 2q\tilde{P}_{(4,1)}(t) + 2q\tilde{P}_{(4,2)}(t) + 2p\tilde{P}_{(2,0)}(t) \\
\qquad\qquad \vdots \\
(m+1)\tilde{P}_{(m+1,1)}(t+1) = mq\tilde{P}_{(m+2,1)}(t) + 2q\tilde{P}_{(m+2,2)}(t) + mp\tilde{P}_{(m,0)}(t) \\
(m+2)\tilde{P}_{(m+2,1)}(t+1) = (m+1)q\tilde{P}_{(m+3,1)}(t) + 2q\tilde{P}_{(m+3,2)}(t) + mp\tilde{P}_{(m+1,0)}(t) + p\tilde{P}_{(m+1,1)}(t) \\
\qquad\qquad \vdots \\
n\tilde{P}_{(n,1)}(t+1) = (n-1)q\tilde{P}_{(n+1,1)}(t) + 2q\tilde{P}_{(n+1,2)}(t) + mp\tilde{P}_{(n-1,0)}(t) + (n-m-1)p\tilde{P}_{(n-1,1)}(t) \\
\qquad\qquad \vdots
\end{cases} \tag{11}
$$

$$\vdots$$

$$
\begin{cases}
m\tilde{P}_{(m,m-1)}(t+1) = q\tilde{P}_{(m+1,m-1)}(t) + mq\tilde{P}_{(m+1,m)}(t) + (m-1)p\tilde{P}_{(m-1,m-2)}(t) + p\left[\sum_{i=0}^{m-2}\tilde{P}_{(m-1,i)}(t)\right] \\
(m+1)\tilde{P}_{(m+1,m-1)}(t+1) = 2q\tilde{P}_{(m+2,m-1)}(t) + mq\tilde{P}_{(m+2,m)}(t) + mp\tilde{P}_{(m,m-2)}(t) \\
(m+2)\tilde{P}_{(m+2,m-1)}(t+1) = 3q\tilde{P}_{(m+3,m-1)}(t) + mq\tilde{P}_{(m+3,m)}(t) + mp\tilde{P}_{(m+1,m-2)}(t) \\
\qquad\qquad + (m+2-m-1)p\tilde{P}_{(m+1,m-1)}(t) \\
\qquad\qquad \vdots \\
n\tilde{P}_{(n,m-1)}(t+1) = (n-m+1)q\tilde{P}_{(n+1,m-1)}(t) + mq\tilde{P}_{(n+1,m)}(t) + mp\tilde{P}_{(n-1,m-2)}(t) \\
\qquad\qquad + (n-m-1)p\tilde{P}_{(n-1,m-1)}(t) \\
\qquad\qquad \vdots
\end{cases} \tag{12}
$$

$$
\begin{cases}
(m+1)\tilde{P}_{(m+1,m)}(t+1) = q\tilde{P}_{(m+2,m)}(t) + (m+1)q\tilde{P}_{(m+2,m+1)}(t) + mp\tilde{P}_{(m,m-1)}(t) \\
\qquad\qquad + p\left[\sum_{i=0}^{m-1}\tilde{P}_{(m,i)}(t)\right] \\
(m+2)\tilde{P}_{(m+2,m)}(t+1) = 2q\tilde{P}_{(m+3,m)}(t) + (m+1)q\tilde{P}_{(m+3,m+1)}(t) + mp\tilde{P}_{(m+1,m-1)}(t) \\
\qquad\qquad + p\tilde{P}_{(m+1,m)}(t) + p\left[\sum_{i=0}^{m}\tilde{P}_{(m+1,i)}(t)\right] \\
\qquad\qquad \vdots \\
n\tilde{P}_{(n,m)}(t+1) = (n-m)q\tilde{P}_{(n+1,m)}(t) + (m+1)q\tilde{P}_{(n+1,m+1)}(t) + mp\tilde{P}_{(n-1,m-1)}(t) \\
\qquad\qquad + (n-m-1)p\tilde{P}_{(n-1,m)}(t) + p\left[\sum_{i=0}^{n-2}\tilde{P}_{(n-1,i)}(t)\right] \\
\qquad\qquad \vdots
\end{cases} \tag{13}
$$

$$\begin{cases} (r+1)\tilde{P}_{(r+1,r)}(t+1) = q\tilde{P}_{(r+2,r)}(t) + (r+1)q\tilde{P}_{(r+2,r+1)}(t) + mp\tilde{P}_{(r,r-1)}(t) \\ (r+2)\tilde{P}_{(r+2,r)}(t+1) = 2q\tilde{P}_{(r+3,r)}(t) + (r+1)q\tilde{P}_{(r+3,r+1)}(t) + mp\tilde{P}_{(r+1,r-1)}(t) \\ \qquad + (r+2-m-1)p\tilde{P}_{(r+1,r)}(t) \\ \qquad \vdots \qquad\qquad\qquad\qquad r \geq m+1 \\ n\tilde{P}_{(n,r)}(t+1) = (n-r)q\tilde{P}_{(n+1,r)}(t) + (r+1)q\tilde{P}_{(n+1,r+1)}(t) + mp\tilde{P}_{(n-1,r-1)}(t) \\ \qquad + (n-m-1)p\tilde{P}_{(n-1,r)}(t) \\ \qquad \vdots \end{cases} \tag{14}$$

Let $K$ be the steady state degree distribution [26-28], and $\Pi(k)$ be the probability distribution of $K$, that is,

$$\Pi(k) = P\{K = k\} = \lim_{t\to+\infty} \sum_{i\geq k+1} \tilde{P}_{(i,k)}(t) \tag{15}$$

Combining Eq.(15) with the state transformation Eqs.(10-14), we can obtain the degree distribution equations of RBDN for different $p$.

**Case 1:** $\mathbf{0 < p < \frac{1}{2}}$

In this case, making use of Eq.(4), we have

$$\lim_{t\to+\infty} \sum_{k\leq n-1} \tilde{P}_{(n,k)}(t) = \Pi_N(n) = \frac{q-p}{q}\left(\frac{p}{q}\right)^{n-1} \tag{16}$$

So combining the state transformation Eqs.(10-14), the degree distribution equations of RBDN can be written as

$$\begin{cases} (q+mp)\Pi(0) = q\Pi(1) + q\Pi_{(1,0)} + \sum_{i=1}^{m-1}(m-i)p\Pi_{(i,0)} \\ (2q+mp)\Pi(1) = 2q\Pi(2) + mp\Pi(0) + p\Pi_{(1,0)} - \sum_{i=1}^{m-1}(m-i)p\Pi_{(i,0)} + \sum_{i=2}^{m-1}(m-i)p\Pi_{(i,1)} \\ \qquad \vdots \\ (mq+mp)\Pi(m-1) = mq\Pi(m) + mp\Pi(m-2) + p\sum_{i=0}^{m-3}\Pi_{(m-1,i)} \\ \left[(m+1)q+mp\right]\Pi(m) = (m+1)q\Pi(m+1) + mp\Pi(m-1) + p - p\sum_{i=1}^{m-1}\Pi_N(i) \\ \left[(m+2)q+mp\right]\Pi(m+1) = (m+2)q\Pi(m+2) + mp\Pi(m) \\ \qquad \vdots \\ \left[(r+1)q+mp\right]\Pi(r) = (r+1)q\Pi(r+1) + mp\Pi(r-1) \\ \qquad \vdots \end{cases} \tag{17}$$

where

$$\Pi_{(i,k)} = \lim_{t\to+\infty} P\left\{NK(t)=(i,k)\right\} \tag{18}$$

For $m=1$ and $m=2$ , we can use the probability generating function method to directly calculate the degree distributions of RBDN.

In the case $m=1$, Eq.(17) can be simplified to

$$\begin{cases} q\left[\Pi(1)-\Pi(0)\right]=p\left[\Pi(0)-\dfrac{q}{p}\Pi_{(1,0)}\right] \\ 2q\left[\Pi(2)-\Pi(1)\right]=p\left[\Pi(1)-\Pi(0)\right]-p \\ 3q\left[\Pi(3)-\Pi(2)\right]=p\left[\Pi(2)-\Pi(1)\right] \\ \qquad\vdots \\ rq\left[\Pi(r)-\Pi(r-1)\right]=p\left[\Pi(r-1)-\Pi(r-2)\right] \\ \qquad\vdots \end{cases} \tag{19}$$

Let the probability generating function be

$$G(x)=\sum_{i=0}^{+\infty}\Pi(i)\cdot x^i,\quad G(1)=\sum_{i=0}^{+\infty}\Pi(i)=1 \tag{20}$$

from Eq. (19), we have

$$G'(x)=\left[\frac{1-px}{q(1-x)}\right]G(x)+\frac{p}{q}-\frac{1}{1-x} \tag{21}$$

Solving Eq. (21), we obtain

$$G(x)=\frac{q}{p}\left(2-\frac{p}{q}-2e^{-p/q}\right)+\frac{q}{p}\sum_{i=1}^{+\infty}x^i\left[2e^{-p/q}\sum_{j=i+1}^{+\infty}\frac{1}{i!}\left(\frac{p}{q}\right)^i\right] \tag{22}$$

Thus for $m=1$, the degree distribution of RBDN is

$$\Pi(k)=\begin{cases} \dfrac{2q}{p}e^{-p/q}\displaystyle\sum_{i=k+1}^{+\infty}\frac{1}{i!}\left(\frac{p}{q}\right)^i-1, & k=0 \\ \dfrac{2q}{p}e^{-p/q}\displaystyle\sum_{i=k+1}^{+\infty}\frac{1}{i!}\left(\frac{p}{q}\right)^i, & k\geq 1 \end{cases} \tag{23}$$

In the case $m=2$ , the degree distribution equations can be simplified to

$$\begin{cases} (q+2p)\Pi(0)=q\Pi(1)+\Pi_{(1,0)} \\ (2q+2p)\Pi(1)=2q\Pi(2)+2p\Pi(0) \\ (3q+2p)\Pi(2)=3q\Pi(3)+2p\Pi(1)+p\left(1-\Pi_{(1,0)}\right) \\ (4q+2p)\Pi(3)=4q\Pi(4)+2p\Pi(2) \\ \qquad\vdots \\ \left[(r+1)q+2p\right]\Pi(r)=(r+1)q\Pi(r+1)+2p\Pi(r-1) \\ \qquad\vdots \end{cases} \tag{24}$$

Using the same method as for $m=1$, the degree distribution of RBDN for $m=2$ is

$$\Pi(k)=\begin{cases}\frac{2+q}{4p}e^{-2p/q}\sum_{i=k+1}^{+\infty}\frac{1}{i!}\left(\frac{2p}{q}\right)^i-\frac{1}{2q}, & k=0\\ \frac{2+q}{4p}e^{-2p/q}\sum_{i=k+1}^{+\infty}\frac{1}{i!}\left(\frac{2p}{q}\right)^i-\frac{p}{2q}, & k=1\\ \frac{2+q}{4p}e^{-2p/q}\sum_{i=k+1}^{+\infty}\frac{1}{i!}\left(\frac{2p}{q}\right)^i, & k\geq 2\end{cases} \tag{25}$$

Different from $m=1$ and $m=2$, for $m\geq 3$, from Eq.(17), we can find that it is necessary to obtain $\Pi_{(i,k)}$ before calculating the degree distribution of RBDN, in which other methods are needed.

**Case 2: $\frac{1}{2}<p<1$**

In this case, making use of Eq.(5), the degree distribution equations of RBDN only determined by the last items of the state transformation Eqs.(10-14). Thus the degree distribution equations can be written as

$$\begin{cases}\left[(m+1)p\right]\Pi(0)=q\Pi(1)\\ \left[q+(m+1)p\right]\Pi(1)=2q\Pi(2)+mp\Pi(0)\\ \vdots\\ \left[(m-1)q+(m+1)p\right]\Pi(m-1)=mq\Pi(m)+mp\Pi(m-2)\\ \left[mq+(m+1)p\right]\Pi(m)=(m+1)q\Pi(m+1)+mp\Pi(m-1)+p\\ \left[(m+1)q+(m+1)p\right]\Pi(m+1)=(m+2)q\Pi(m+2)+mp\Pi(m)\\ \vdots\\ \left[rq+(m+1)p\right]\Pi(r)=(r+1)q\Pi(r+1)+mp\Pi(r-1)\\ \vdots\end{cases} \tag{26}$$

Constructing the probability generating function $G(x)=\sum_{i=0}^{+\infty}\Pi(i)*x^i, G(1)=1$, from Eq. (26), we have

$$G'(x)=\left[\frac{mp}{q}+\frac{p}{q(1-x)}\right]G(x)-\frac{p}{q}\frac{x^m}{1-x} \tag{27}$$

Solving Eq. (27), we get

$$G(x)=\frac{ce^{mcx}}{(1-x)^c}\int_x^1 t^m(1-t)^{c-1}e^{-mct}dt \tag{28}$$

where $c=\frac{p}{q}>1$. Let

$$1 - t = y \tag{29}$$

so we have

$$\begin{aligned}
G(x) &= \frac{ce^{mcx}}{(1-x)^c}\int_0^{1-x}(1-y)^m y^{c-1}e^{-mc(1-y)}dy \\
&= \frac{ce^{-mc}e^{mcx}}{(1-x)^c}\sum_{i=0}^{m}C_m^i(-1)^i\int_0^{1-x}y^{i+c-1}e^{mcy}dy \\
&= \frac{ce^{-mc}e^{mcx}}{(1-x)^c}\sum_{i=0}^{m}C_m^i(-1)^i\sum_{j=0}^{+\infty}\frac{(mc)^j}{j!}\int_0^{1-x}y^{i+c-1+j}dy \\
&= ce^{-mc}e^{mcx}\sum_{i=0}^{m}C_m^i(-1)^i\sum_{j=0}^{+\infty}\frac{(mc)^j}{j!}\frac{1}{i+j+c}(1-x)^{i+j} \\
&= ce^{-mc}e^{mcx}\sum_{j=0}^{+\infty}x^j\left[\sum_{n=j}^{+\infty}C_n^j(-1)^j\frac{1}{n+c}\left(\sum_{i=0}^{\min(m,n)}C_m^i(-1)^i\frac{(mc)^{n-i}}{(n-i)!}\right)\right] \\
&= ce^{-mc}\sum_{r=0}^{+\infty}x^r\left[\sum_{j=0}^{r}(-1)^j\sum_{n=j}^{+\infty}C_n^j\frac{1}{n+c}\left(\sum_{i=0}^{\min(m,n)}C_m^i(-1)^i\frac{(mc)^{n-i}}{(n-i)!}\right)\frac{(mc)^{r-j}}{(r-j)!}\right]
\end{aligned} \tag{30}$$

Therefore the degree distribution is

$$\Pi(k) = ce^{-mc}\sum_{j=0}^{k}(-1)^j\sum_{n=j}^{+\infty}C_n^j\frac{1}{n+c}\left(\sum_{i=0}^{\min(m,n)}C_m^i(-1)^i\frac{(mc)^{n-i}}{(n-i)!}\right)\frac{(mc)^{k-j}}{(k-j)!} \qquad k = 0,1,2,\cdots \tag{31}$$

**Case 3:** $\boldsymbol{p = \frac{1}{2}}$

In this case, making use of Eq.(6) and the state transformation Eqs.(10-14). The degree distribution equations can be written as

$$\begin{cases}
(q+mp)\Pi(0) = q\Pi(1) \\
(2q+mp)\Pi(1) = 2q\Pi(2) + mp\Pi(0) \\
\qquad\vdots \\
(mq+mp)\Pi(m-1) = mq\Pi(m) + mp\Pi(m-2) \\
\left[(m+1)q+mp\right]\Pi(m) = (m+1)q\Pi(m+1) + mp\Pi(m-1) + p \\
\left[(m+2)q+mp\right]\Pi(m+1) = (m+2)q\Pi(m+2) + mp\Pi(m) \\
\qquad\vdots \\
\left[(r+1)q+mp\right]\Pi(r) = (r+1)q\Pi(r+1) + mp\Pi(r-1) \\
\qquad\vdots
\end{cases} \tag{32}$$

We may use the probability generating function approach or the recursive method to solve the Eq.(32). The result is same as that in Ref. [24].

$$\Pi(k)=\begin{cases}(m-1)!m^{-m}e^{-m}\sum_{j=m+1}^{+\infty}\frac{m^j}{j!}\cdot\sum_{r=0}^{k}\frac{m^r}{r!} & 0\le k\le m\\(m-1)!m^{-m}e^{-m}\sum_{j=0}^{m}\frac{m^j}{j!}\cdot\sum_{r=k+1}^{+\infty}\frac{m^r}{r!} & k\ge m+1\end{cases} \tag{33}$$

## 5. Simulation and Tail Characteristics

Before discussing the tail characteristics of the degree distribution for RBDN, it is necessary to verify our theoretical results in Section 4. We do this by computer simulation. Figures 1-3 illustrate the exact solutions and simulation results of the degree distributions for different $p$, where the horizontal and vertical ordinates denote the degree of nodes and the probability respectively. Each simulation number is the average value of 1000 simulation results for $t=10000$. As shown in Fig. 1 and 2, in the case of $0<p<\frac{1}{2}$, $m=1,2$, the exact solutions match perfectly with the simulation results and the correctness of our exact solutions can be verified.

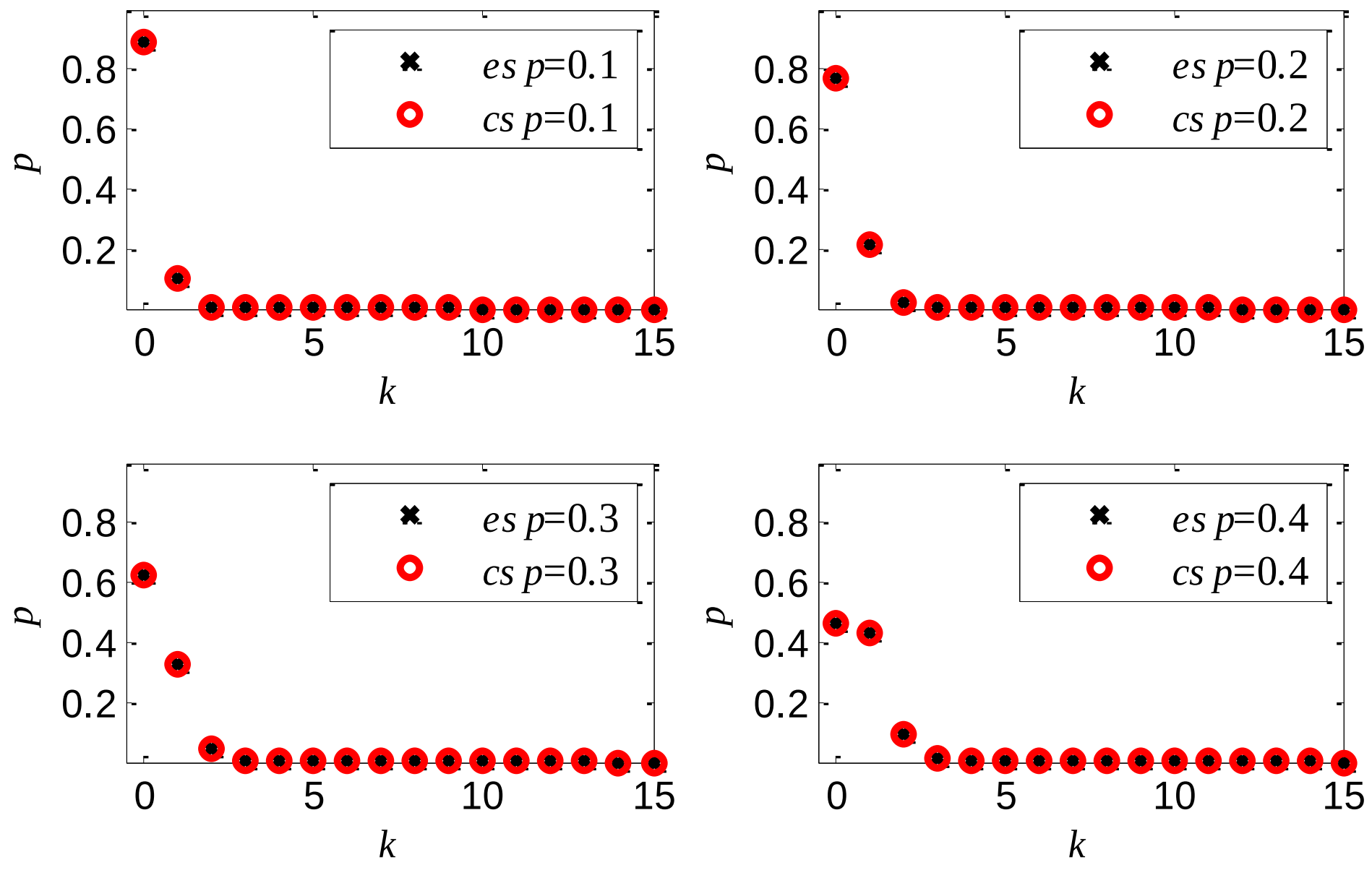

Fig. 1 Exact solutions (*es*) vs. computer simulation (*cs*): the degree distributions of RBDN for $0<p<\frac{1}{2}$, $m=1$ (All statistical errors are smaller than $10^{-3}$)

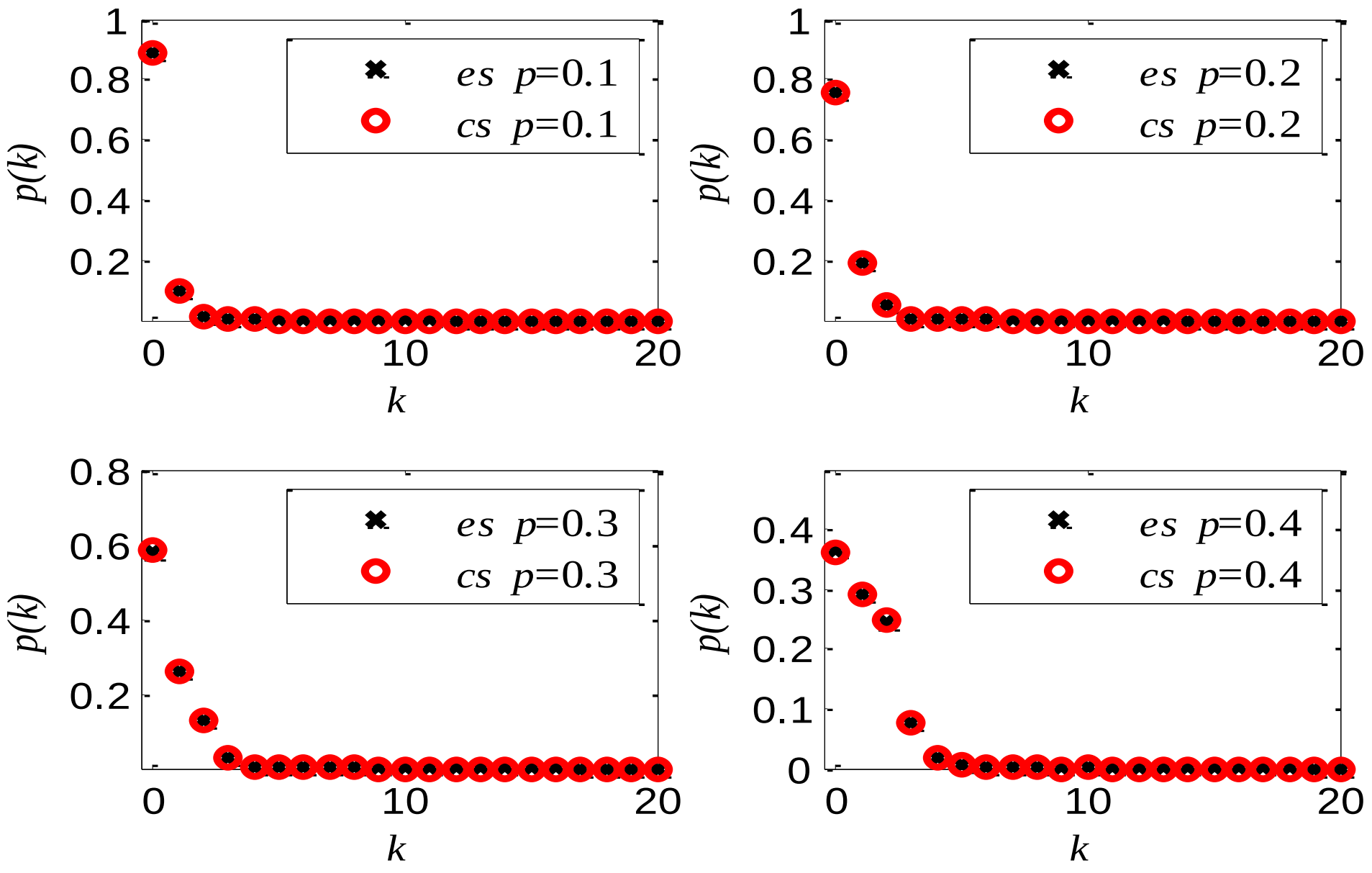

Fig.2 Exact solutions (*es*) vs. computer simulation (*cs*): the degree distributions of RBDN

for $0 < p < \frac{1}{2}, \ m = 2$ (All statistical errors are smaller than $10^{-3}$)

Figures 3 provides the comparisons of exact solutions and computer simulation for the degree distributions of RBDN in the case of $\frac{1}{2} < p < 1, \ m = 3$. We may find that the simulation results match exact solutions very well.

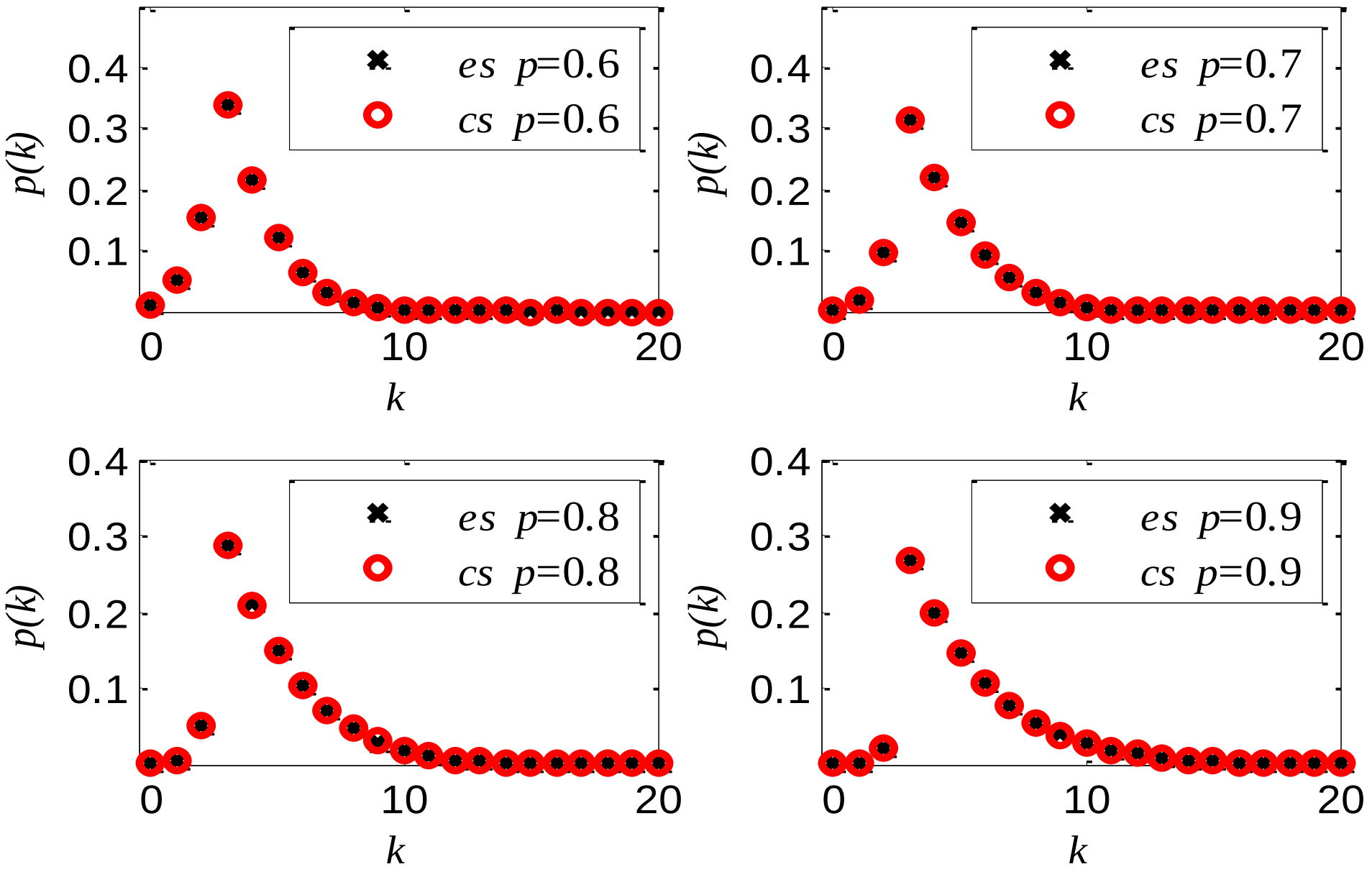

Fig. 3 Exact solutions (*es*) vs. computer simulation (*cs*)： the degree distributions of RBDN

for $\frac{1}{2} < p < 1, \ \ m = 3$ (All statistical errors are smaller than $10^{-3}$)

The tail characteristics of the degree distribution are important topics for complex networks. Currently, the literature on degree distribution discovers the power-law tail for a scale-free network [26], the Poisson tail for a small-world network [32], and the exponential tail for a growing exponential network [8].

For the large $k\left(k \geq m+1\right)$, let

$$r\left(k\right)=\frac{\Pi\left(k\right)}{\Pi\left(k-1\right)} \tag{34}$$

If $\Pi\left(k\right)$ approximately follows a Poisson distribution, i.e.

$$\Pi\left(k\right) \sim \alpha\frac{\lambda^{k}}{k!} \tag{35}$$

then we have

$$r\left(k\right) \sim \frac{\lambda}{k}, \quad \ln r\left(k\right) \sim -\ln k \tag{36}$$

In other words, if $\Pi\left(k\right)$ is subjected to a Poisson distribution, then the relationship of $\ln r\left(k\right)$ and $\ln k$ in the dual-logarithm coordinate system should be a straight line with slope of -1.

If $\Pi\left(k\right)$ is nearly subjected to an exponential distribution, i.e.

$$\Pi\left(k\right) \sim e^{-\alpha k} \tag{37}$$

then

$$r\left(k\right) \sim e^{-\alpha} \tag{38}$$

In other words, if $\Pi\left(k\right)$ approximately follows an exponential distribution, the relationship of $\ln r\left(k\right)$ and $\ln k$ in the dual-logarithm coordinate system should be a straight line with slope of 0. Thus we may further explore the tail characteristics of the degree distribution of RBDN.

Fig. 4 illustrates the tails of the degree distribution for RBDN in the case of $0 < p \leq \frac{1}{2}, m=2$. As shown in Fig.4, for different $p$, when $k \geq 10$, the slopes of lines tend to be -1. So $\Pi\left(k\right)$ exhibits Poisson tail.

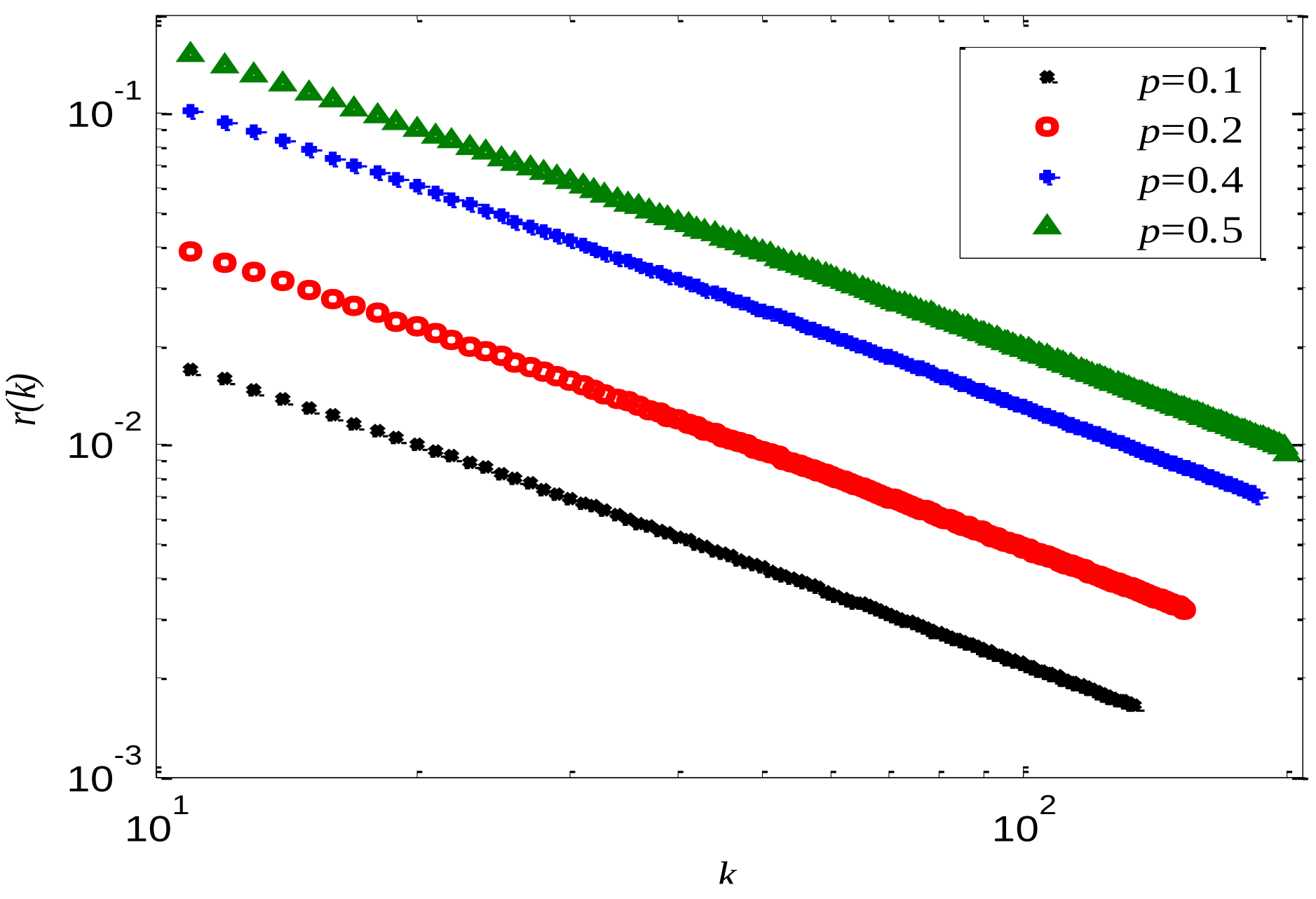

Fig. 4 The tails of the degree distribution of RBDN: $p \le \frac{1}{2}$, $m = 2$ (Poisson tail)

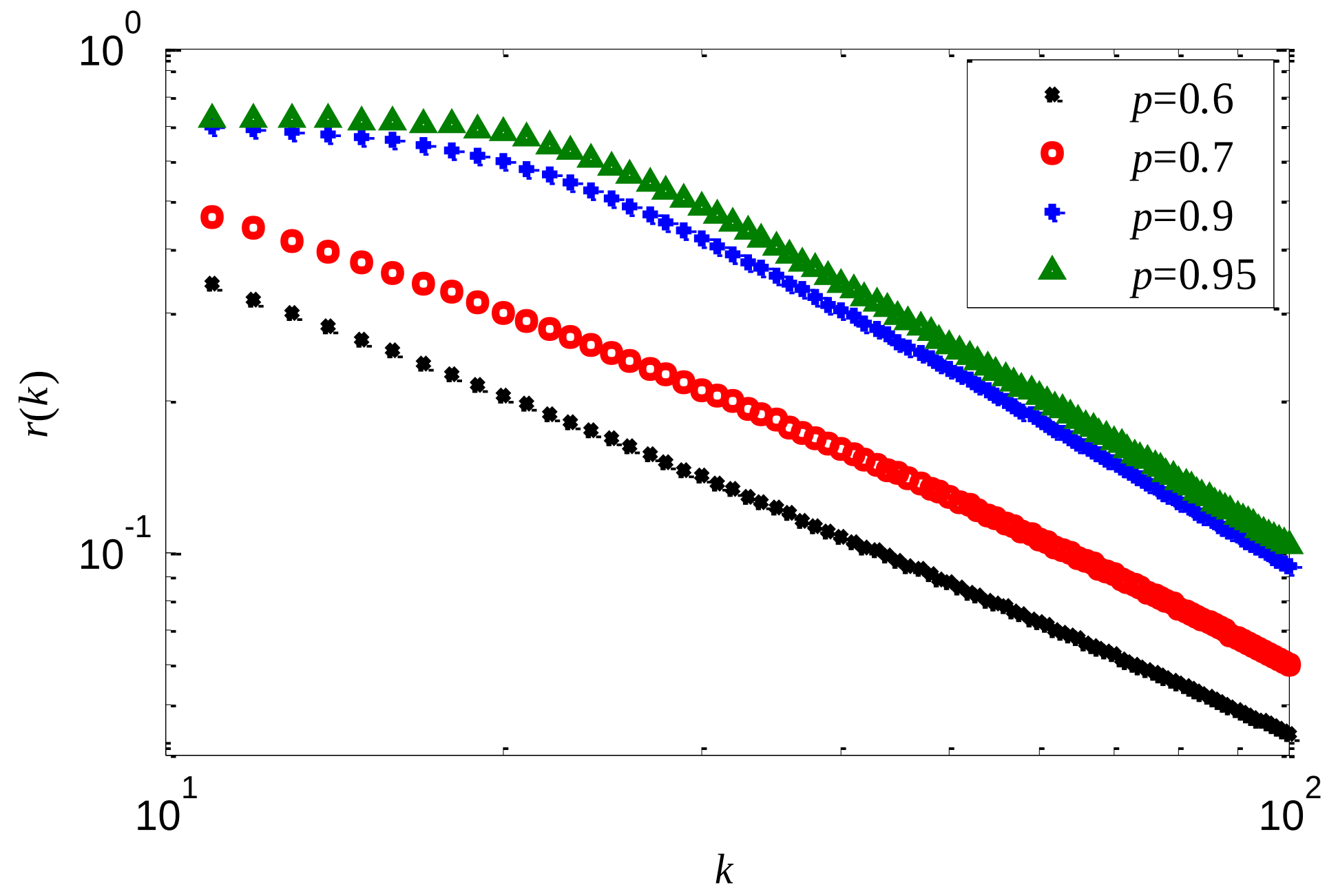

Fig. 5 The tails of the degree distribution of RBDN: $\frac{1}{2} < p$, $m = 3$ (From Poisson tail to exponential tail)

In the case of $\frac{1}{2} < p < 1$, as illustrated in Fig.5, the tail of the degree distribution also changes with $p$, exhibiting Poisson tail in the case $p$ close to $\frac{1}{2}$ (e.g. $p = 0.7, p = 0.6$). However with the increasing of $p$, Poisson tails disappear gradually. In the case $p = 0.95$, for $10 < k < 20$, $r(k)$ tends to be a constant (for $p = 1$, $r(k)$ is a constant.), showing the tail of the degree distribution approximately exhibits an exponential tail as $p \to 1$.

## 6. Conclusion

In this paper, a random birth-and-death network (RBDN) model is introduced, in which at each time step, a new node is added with probability $p$ or an old node is removed from the network with probability $q = 1 - p$. As a first step in investigating the property of RBDN, we have explored the network size, the degree distribution and its tail characteristics. Exact solutions of degree distribution in the case of $\frac{1}{2} < p < 1$ and $0 < p < \frac{1}{2}$ have been calculated and compared with simulation results. We find that the tail of the degree distribution for RBDN exhibits a Poisson tail in the case of $0 < p \le \frac{1}{2}$ and an exponential tail as $p$ approaches to 1.

For RBDN, this prototype investigation could be developed further. First the relationship between connectivity and $p$ should be considered. Second, other properties like average path length and cluster coefficient need to be further explored. Third, the dynamic behavior of RBDN needs further investigation since it may provide deeper understanding of evolving networks whose size varies over time.

## Acknowledgments

This research is financially supported by the National Natural Science Foundation of China (No. 61273015) and the China Scholarship Council.

## Appendix: one-step transition probability matrix *P*

Using SPR, the one-step transition probability matrix $\boldsymbol{P}$ has two possibilities:

1. Add a node

i. To be an isolated node, node $v$ connects to other networks at time $t+1$, then the state of node $v$ turns from $(n,k)$ to $(n+1,m)$ or $(n+1,n)$, and the one-step transition probability is given by

$$p_{(n,k),(n+1,m)} = P\left\{KV(t+1) = (n+1,m) \middle| KV(t) = (n,k)\right\} = \frac{1}{2(n+1)}, \quad n \ge m, 0 \le k < n \tag{39}$$

$$p_{(n,k),(n+1,n)} = P\left\{KV(t+1) = (n+1,n) \middle| KV(t) = (n,k)\right\} = \frac{1}{2(n+1)}, \quad n < m, 0 \le k < n \qquad (40)$$

ii. Node $v$ does not connect to the new added node at time $t+1$, then the state of node $v$ turns from $(n,k)$ to $(n+1,k)$ and one-step transition probability is given by

$$p_{(n,k),(n+1,k)} = P\left\{KV(t+1) = (n+1,k) \middle| KV(t) = (n,k)\right\} = \frac{n-m}{2(n+1)}, \quad n \ge m, 0 \le k < n \qquad (41)$$

iii. Node $v$ connects to the new added node at time $t+1$, then the state of node $v$ turns from $(n,k)$ to $(n+1,k+1)$ and one-step transition probability is given by

$$p_{(n,k),(n+1,k+1)} = P\left\{KV(t+1) = (n+1,k+1) \middle| KV(t) = (n,k)\right\} = \frac{m}{2(n+1)}, \quad n \ge m, 0 \le k < n \qquad (42)$$

$$p_{(n,k),(n+1,k+1)} = P\left\{KV(t+1) = (n+1,k+1) \middle| KV(t) = (n,k)\right\} = \frac{n}{2(n+1)}, \quad n < m, 0 \le k < n \qquad (43)$$

2. <u>Delete a node</u>

Since any node in the network with node $v$ may be deleted with equal probability, we only need to compute the transition probability of nodes not being deleted.

iv. The degree of node $v$ is decreased by 1 at time $t+1$, Then the state of node $v$ turns from $(n,k)$ to $(n-1,k-1)$, and the one-step transition probability is given by

$$p_{(n,k),(n-1,k-1)} = P\left\{KV(t+1) = (n-1,k-1) \middle| KV(t) = (n,k)\right\} = \frac{k}{2(n-1)}, \quad 1 \le k < n \qquad (44)$$

$$p_{(n,0),(n,0)} = P\left\{KV(t+1) = (n,0) \middle| KV(t) = (n,0)\right\} = \frac{1}{2}, \quad n = 1 \qquad (45)$$

v. The degree of node $v$ remains unchanged at time $t+1$, Then the state of node $v$ turns from $(n,k)$ to $(n-1,k)$, and the one-step transition probability is given by

$$p_{(n,k),(n-1,k)} = P\left\{KV(t+1) = (n-1,k) \middle| KV(t) = (n,k)\right\} = \frac{n-1-k}{2(n-1)}, n \ge k+1 \ge 1 \qquad (46)$$